\documentstyle[preprint,aps]{revtex}
\begin{document}
\draft
\preprint{RU9448}
\title
{A Limited Symmetry Found by Comparing Calculated Magnetic
Dipole Spin and Orbital Strengths in $^4\mbox{He}$}

\author
{M.S. Fayache and L. Zamick}

\address
{Department of Physics and Astronomy, Rutgers University,
Piscataway, New Jersey 08855-0849}

\date{today}

\maketitle

\begin{abstract}
Allowing for $2$$\hbar \omega$ admixtures in $^4\mbox{He}$ we find
that the summed magnetic dipole isovector orbital and spin strengths
are equal. This indicates a symmetry which is associated with
interchanging the labels of the spin with those of the orbit. Where
higher admixtures are included, the orbital sum becomes larger than
the spin sum, but the sums over the low energy region are still nearly
the same.
\end{abstract}

\narrowtext

In an $LS$ closed shell nucleus e.g. $^4\mbox{He}$, $^{16}\mbox{O}$,
$^{40}\mbox{Ca}$ the magnetic dipole transition from the $J$=$0^+$
ground state to the $J$=$1^+$ excited states will vanish unless there
are ground state correlations such as $2$-particle $2$-hole
admixtures. Previous theoretical studies of magnetic spin dipole
excitations~\cite{zah} show that the correlations induced by the
tensor interaction give a large contribution to the energy weighted
sum rule in a closed shell nucleus. To see the full effect of the
tensor interaction, one has to allow excitations up to large values
of $n \hbar \omega$. A simplifying feature in the above calculations
is the observation, at least for the isoscalar transitions, that the
double commutator of the isoscalar magnetic dipole spin operator with
the tensor interaction is proportional to the same tensor interaction.
Another point made in the above work was that a central interaction
had to have a spin dependence in order to generate magnetic dipole
strength in a closed $LS$ shell nucleus.

Other works on $M1$'s with sum rule techniques include those of
Desplanques et. al.~\cite{dp}, Orlandini et. al.~\cite{or} and a
Physics Report by Lipparini and Stringari~\cite{st}.

Some experimental and experiment-theory collaborative works for
magnetic dipole transitions in $^{16}\mbox{O}$ and $^{40}\mbox{Ca}$
have been performed with a motivation of discerning the nature of
ground state correlations. These include the inelastic scattering work
of W. Gross et. al.~\cite{gr}, A. Richter et. al.~\cite{rch}, W.
Steffen et. al.~\cite{stf} and B. A. Brown et. al.~\cite{br} and the
proton capture work of R. A. Snover et. al.~\cite{sn} in $^{16}\mbox{O}$.

Whereas in our earlier work~\cite{zah} we considered only spin
transitions, in the present work we wish to consider also
{\underline {orbital}} magnetic dipole transitions and to see if there
is any interrelation with the spin transitions in a closed shell
nucleus. For example, does the tensor interaction also induce
{\underline {orbital}} excitations comparable to the spin excitations $?$

We shall calculate the magnetic dipole strengths from the $J$=$0^+$
$T$=$0$ of $^4\mbox{He}$ to $J$=$1^+$ $T$=$1$ excited states. We will
calculate separately the total $B(M1)$ rate, $B(M1)_{spin}$ and
$B(M1)_{orbit}$ where the operators in question, in units of
${\mu_N}$ are $9.412$$\vec{s}$ $t_z$ + $\vec{l}$ $t_z$, $\vec{s}$
$t_z$ and $\vec{l}$ $t_z$ where $t_z$=$+\frac{1}{2}$ for a proton and
$-\frac{1}{2}$ for a neutron. Note that we define our isovector spin
$B(M1)$ so it has the same coupling strength as the orbital one i.e.
we drop the factor $9.412$. This makes it easier to compare spin and
orbital strengths.

As mentioned previously, it is necessary to have ground state
correlations in $^4\mbox{He}$ in order to get magnetic dipole
transitions. We get these by performing shell model matrix
diagonalizations using OXBASH~\cite{oxbash}. We have used
progressively larger shell model spaces for the $J$=$0^+$ $T$=$0$
ground state and $J$=$1^+$ $T$=$1$ states: up to $2$$\hbar\omega$, up
to $4$$\hbar\omega$, and up to $6$$\hbar\omega$ admixtures.

We use the interaction of Zheng and Zamick~\cite{zz} which has
central, spin-orbit and tensor parts:

\begin{equation}
V_{sche}=V_c+xV_{so}+yV_t
\end{equation}

For $x$=$1$ and $y$=$1$ this interaction gives a fairly good fit to the matrix
elements of the BONN A interaction. We can turn the spin-orbit
(tensor) interaction off by setting $x$ ($y$) equal to zero. We can
thus isolate and study the effects of the spin-orbit and/or tensor
interaction on the magnetic dipole excitations.

In Table $I$ we give the total summed strength $B(M1)_{spin}$ and
$B(M1)_{orbit}$ to all (non-spurious) $J$=$1^+$ $T$=$1$ states
corresponding to the operators $\vec{s}$ $t_z$ and $\vec{l}$ $t_z$,
respectively (as mentioned before we drop the isovector factor
$5.586-(-3.826)$=$9.412$). We do this for progressively increasing
model spaces: up to $2$$\hbar \omega$, up to $4$$\hbar \omega$ and up
to $6$$\hbar \omega$.

We perform the calculation with the spin-orbit and tensor interactions
off and on.

Examining Table $I$ we find one a priori unexpected result.
When we restrict the ground state correlations to $2$$\hbar \omega$,
we find that the summed \underline{spin} strengths are virtually equal
to the summed \underline{orbital} strengths. This is true for all four
cases of
($x$,$y$) i.e. whether or not there is a spin-orbit interaction
present and whether or not there is a tensor interaction present.

This striking result does not extend to larger configurations. When we
allow up to $4$$\hbar \omega$ excitations, the summed orbital $B(M1)$
strengths become substantially larger than the summed spin
strengths. For example, for $x$=$1$ and $y$=$1$, these are
respectively $10.6$ ($10^{-3}{\mu_N}^2)$ and $4.8$ ($10^{-3}{\mu_N}^2)$.
Note that when we go to $6$$\hbar \omega$ excitations, the summed
strengths are even larger, and the convergence in terms of
$n$$\hbar \omega$, if it exists, is very slow.

Let us now consider the systematics of the interaction. We note that,
relative to the central (but spin-dependent) force case ($x$=$0$,
$y$=$0$), turning on the spin-orbit interaction scarcely changes the
summed strength at all. However, when the tensor interaction is turned
on (by changing $y$ from $0$ to $1$), there is a big jump in the
summed spin strength and in the summed orbital strength. In the
$2$$\hbar \omega$ case, the change in both cases (since they are
equal) is from $0.86$ ($10^{-3}{\mu_N}^2$) to $3.82$ ($10^{-3}{\mu_N}^2$).

We gain further insight by examining Table II where the strengths to
individual states are given.
Consider first the $2$$\hbar \omega$ calculation -and the case
of a central interaction ($x$=$0$, $y$=$0$). There are only seven
non-spurious $J$=$1^+$ $T$=$1$ states in this model space with the
following excitation energies in $MeV$: $36.7$, $44.0$, $45.3$,
$48.8$, $49.2$, $53.9$ and $56.4$. The spin transition strength goes
to only one state -the third one at $45.3$ $MeV$ with a strength
$B(M1)_{spin}$=$0.8546$ ($10^{-3}{\mu_N}^2$). The orbital transition
strength also goes to one state, but to a different one than in the
case of the spin. The orbital strength all goes to the highest state
(\#$7$) at $56.5$ $MeV$ with a value
$B(M1)_{orbit}$=$0.8546$ ($10^{-3}{\mu_N}^2$). Thus
$B(M1)_{orbit}$=$B(M1)_{spin}$. For other interactions $x$$\neq$$0$
or $y$$\neq$$0$, we don't get these sharp results although one can
make an approximate association between nearly equal spin and orbital
transitions. Nevertheless, the summed values of $B(M1)_{spin}$ and
$B(M1)_{orbit}$ are virtually the same even in the presence of
spin-orbit and tensor interactions.

The above behaviour more or less tells us what the symmetry we are
dealing with is.
For a central interaction, given one eigenstate with certain spin and
orbital labels, we can get another eigenstate by interchanging the
spin labels with the orbital labels. This symmetry is limited to the
$2$$\hbar \omega$ space because in that space the configurations are
simple -all are of the form $0s^2$$0p^2$.

We consider the case $x$=$y$=$0$. We are in the $L$$S$ limit. Since
the $0s^4$ closed shell has $L$=$0$ $S$=$0$, only $2$$\hbar \omega$
excitations with the same quantum numbers will admix into the ground
state. Let us consider two particles excited from the $0s$ shell to
the $0p$ shell. We can label the $2p$-$2h$ states by
${\lbrack L_{\pi} L_{\nu} \rbrack}^{L=0}{\lbrack S_{\pi} S_{\nu}
\rbrack}^{S=0}$ . There are several cases to be considered:

1. Two protons are excited. The configurations are
$(p_{\pi}^2)^{L_{\pi}S_{\pi}}(s_{\nu}^2)^{L_{\nu}S_{\nu}}$. Since
$L_{\nu}$=$0$ and $S_{\nu}$=$0$ and $L$ and $S$ are zero, we must have
$L_{\pi}$=$0$ and $S_{\pi}$=$0$. So all in all we get the state :
$|a\rangle=(p^2)^{L_{\pi}=0,S_{\pi}=0}(s^2)^{L_{\nu}=0,S_{\nu}=0}$.

2. Two neutrons are excited. By analogy, the configuration is
$|b\rangle=(s^2)^{L_{\pi}=0,S_{\pi}=0}(p^2)^{L_{\nu}=0,S_{\nu}=0}$.

3. A neutron and a proton are excited from the $s$ shell to the $p$
shell. The configuration is :
${\lbrack(sp)^{L_{\pi}S_{\pi}}(sp)^{L_{\nu}S_{\nu}}\rbrack}^{L=0,
S=0}$. There are two possibilities:

$|c\rangle={\lbrack{L_{\pi}=1,
L_{\nu}=1}\rbrack}^{L=0}{\lbrack{S_{\pi}=0, S_{\nu}=0}\rbrack}^{S=0}$
and

$|d\rangle={\lbrack{L_{\pi}=1,
L_{\nu}=1}\rbrack}^{L=0}{\lbrack{S_{\pi}=1, S_{\nu}=1}\rbrack}^{S=0}$.

We can form an isovector orbital excitation by applying the operator
$\vec{L_{\pi}}-\vec{L_{\nu}}$ to the $J$=$0^+$ ground state; likewise
we can form an isovector spin excitation by applying the operator
$\vec{S_{\pi}}-\vec{S_{\nu}}$ to the $J$=$0^+$ ground state.
When acting on the configurations $|a\rangle$ or $|b\rangle$, the
orbital operator $\vec{L_{\pi}}-\vec{L_{\nu}}$ gives zero; likewise
the spin operator $\vec{S_{\pi}}-\vec{S_{\nu}}$. That is:

$(\vec{L_{\pi}}-\vec{L_{\nu}})|L_{\pi}=0, L_{\nu}=0\rangle=0$

Let us skip to the state $|d\rangle$. Note that the orbital and spin
quantum numbers are the same: $L_{\pi}=S_{\pi}=1$ and
$L_{\nu}=S_{\nu}=1$. This is enough to prove that, if this were the
only state present, we would have the result:
$B(M1)_{spin}$=$B(M1)_{orbit}$.

In more detail,
$$(\vec{L_{\pi}}-\vec{L_{\nu}})|d\rangle=N{\lbrack{L_{\pi}=1,
L_{\nu}=1}\rbrack}^{L=1}{\lbrack{S_{\pi}=1, S_{\nu}=1}\rbrack}^{S=0}$$
and
$$(\vec{S_{\pi}}-\vec{S_{\nu}})|d\rangle=N{\lbrack{L_{\pi}=1,
L_{\nu}=1}\rbrack}^{L=0}{\lbrack{S_{\pi}=1, S_{\nu}=1}\rbrack}^{S=1}$$

There is no reason why these states should be at the same energy and
indeed they are not, but the equality of the spin and orbital
strengths, {\underline {provided}} the state $|c\rangle$ were not
present, is obvious. However, the presence of the state $|c\rangle$
apparently presents a problem. The isovector spin operator
$\vec{S_{\pi}}-\vec{S_{\nu}}$ will annihilate this state, whereas the
isovector orbital operator ($\vec{L}_{\pi} - \vec{L}_{\nu}$) creates
the state ${\lbrack{L_{\pi}=1,
L_{\nu}=1}\rbrack}^{L=1}{\lbrack{S_{\pi}=0, S_{\nu}=0}\rbrack}^{S=0}$.
There should therefore be more orbital strength than spin strength.
What saves the day is that this transition is spurious. In the OXBASH
program \cite{oxbash} the spurious states are put very high in energy
by adding a large constant to the single particle energies for center
of mass motion. We added $100$ $MeV$ for each nucleon thus putting the
spurious states in the vicinity of $400$ $MeV$ excitation energy. In
table $III$ we show the $2$$\hbar\omega$ $x$=$0$ $y$=$0$ calculation
in which all the $1^+$ $T$=$1$ states are shown, both non-spurious and
spurious, with the values of $B(M1)_{spin}$ and $B(M1)_{orbit}$.

We see from Table II that our results are consistent with the above
discussion. The
non-spurious orbital and spin strengths are the same:
$0.855$ ($10^{-3}{\mu_N}^2$), but the respective states are at different
energies $45.3$ $MeV$ for the spin state and $56.5$ $MeV$ for the
orbital state. These correspond to excitations from the configuration
$|d\rangle$. There are no more spin excitations but there is an
orbital excitation which is quite strong $13.07$ ($10^{-3}{\mu_N}^2$) to
a spurious state artificially placed at $439.3$ $MeV$. This is
consistent with our previous remarks that the configuration
$|c\rangle$ allows for an orbital but not a spin excitation.

We can further extend our results to include the tensor interaction.
This interaction allows $[L=2~ S=2]^{J=0}$ 2-particle-2 hole admixtures into
the ground state. For 2$\hbar \omega$ excitations the only way to
achieve such a state is to excite a proton and neutron from the $0s$
state to the $0p$  state. Thus we have a state
$$ |e > ~=~ \{ [L_{\pi} = 1, L_{\nu} = 1]^{L=2} ~~ [S_{\pi} = 1 , S_{\nu}
= 1 ]^{S=2}\}^{J=0}$$
The state $|e>$ is also invariant under the interchange of the spin
and orbit labels, and hence preserves the equality of the summed spin
and summed orbit strength at the 2$\hbar \omega$ level.

Concerning the two body spin-orbit interaction it should be noted that
all matrix elements of the form $< 0s ~ 0s  V_{s.o.} 0p ~ 0p>$  vanish.
The reason is that the spin orbit interaction does not act as in
relative $\ell$ = 0 states and furthermore does not change the
relative orbital angular momentum. However the $0s$  $0s$  state can only
have relative orbital angular momentum equal to zero. Thus the spin
orbit interaction does not induce ground state correlations in first
order perturbation theory.

We thus have explained the equality of the summed spin and summed
orbital strength in $^4$He for the entire interaction - central,
spin-orbit and tensor. It should be noted these results are specific
to $^4$He.
For larger closed shell nuclei e.g.
$^{16}\mbox{O}$, the orbital $B(M1)$ is substantially larger than the
spin $B(M1)$ even at the $2$ $\hbar\omega$ level \cite{ba}.

It is trivial to show that for an isoscalar magnetic dipole transition
from the $J$=$0^+$ $T$=$0$ ground state to a given $J$=$1^+$ $T$=$0$
excited state, the matrix element of the spin operator $\vec{S}$=
$\vec{S_{\pi}}+\vec{S_{\nu}}$ is equal and opposite to that of the
orbital operator $\vec{L}$=$\vec{L_{\pi}}+\vec{L_{\nu}}$. This is
because the total angular momentum operator $\vec{J}$=
$\vec{L}+\vec{S}$, when acting on the $J$=$0^+$ ground state yields
zero. More generally, since any nuclear state is an eigenfunction of
$\vec{J}$, this operator cannot induce transitions out of the
multiplet.

However, the above argument certainly does not hold for the isovector
case for which the relevant operators are
$\vec{L}$=$\vec{L_{\pi}}-\vec{L_{\nu}}$ and
$\vec{S}$=$\vec{S_{\pi}}-\vec{S_{\nu}}$. Furthermore the equality that
we obtain between spin and orbit in the isovector case (at the
$2$$\hbar\omega$ level) is for different states, whereas in the
isoscalar case it is for the same $1^+$ state. It should be further
noted that one does not get any isoscalar $1^+$ transitions in an $LS$
closed shell like $^4\mbox{He}$ in the case of a central
spin-dependent interaction \cite{zah}. However, if a tensor
interaction is present, we do get finite isoscalar transitions.

In the case $x$=$0$ $y$=$0$ when we allow up to $4$ $\hbar\omega$ or
$6$ $\hbar\omega$ excitations, we no longer have the summed orbital
strengths equal. However, some features of the $2$ $\hbar\omega$ case
are preserved in the $6$ $\hbar\omega$ calculation. Most transition
rates vanish. In the low energy sector (defined more precisely in the
next section) only one spin state and only
one orbital state get excited, just as in the $2$ $\hbar\omega$ case. The
spin state is at $34.4$ $MeV$ with
$B(M1)$=$0.55$ ($10^{-3}{\mu_N}^2$) and the orbital state is at
$43.5$$MeV$ with $B(M1)$=$0.60$ ($10^{-3}{\mu_N}^2$). Although the two
$B(M1)'s$ are not equal they differ by less than $11$$\%$, as shown in
Table $III$.

But other states in the $4$ $\hbar\omega$ and $6$ $\hbar\omega$ region
also get excited. Indeed, the single largest calculated orbital
$B(M1)$ is to a state at $61.4$$MeV$ with a rate
$B(M1)_{orbit}$=$2.79$ ($10^{-3}{\mu_N}^2$). This is more than four
times larger than the $B(M1)$ in the low energy sector. We show in
Table IV for $x$=$0$ $y$=$0$ all states with $B(M1)$ $\ge$
$10^{-4}{\mu_N}^2$ .

In Fig.~1 we present the cumulative sum of the strength
distribution for the spin $B(M1)$ and orbit $B(M1)$ when up to $6$
$\hbar\omega$ are allowed. We consider the case $x$=$1$, $y$=$1$
(realistic). In Fig.~1 we give the spin distribution. We see some
strength starting at about $35$ $MeV$ with a plateau from about
$41$$MeV$  until about $65$$MeV$. This is the low lying strength which
one might obtain in a $2$$\hbar\omega$ restricted space.
Then there is a sharp rise corresponding to $4$ $\hbar\omega$ and
higher admixtures. The curve ultimately flattens out because we run
out of states. The corresponding orbital strength curve also has a
plateau from about $46$ to $61$ $MeV$. This also can be identified as
the low energy part. As mentioned in the previous section, the value of
$B(M1)_{orbit}$ is very close to the value of $B(M1)_{spin}$ for this
plateau. This shows that the symmetry relation for $2$$\hbar\omega$ is
not broken very much in the low energy sector when we extend the the
calculation to $6$$\hbar\omega$. The low energy strength is obviously
easier to find experimentally than the higher lying strength.

As we increase the excitation energy in the orbital case, we see a
sharp rise at $63$ $MeV$ to another  plateau. The second orbital
plateau is much higher than the second spin plateau. But then, unlike
the spin case, there are more sharp rises until we reach a saturation
value of $14.3$ ($10^{-3}{\mu_N}^2$). It would certainly be of
interest to look for such a strong orbital strength distribution at a
very high energy $\sim$ $3$ to $4$ $\hbar\omega$.
If we had extended our
calculation to $8$ $\hbar\omega$ there might be even further rises.

In closing we point out that we have uncovered a rather unusual
symmetry when $2$ $\hbar\omega$ ground state correlations are included
in the wave function of $^4\mbox{He}$. It will be difficult to test
this result experimentally because of the large isovector spin
coupling for the electromagnetic probe which will drown out the
orbital contribution. Possibly, a multi-probe  analysis would help.
Nevertheless, we feel that the results are of considerable {\underline
{theoretical}} interest. Among the unique features of our findings are:

(a) We obtain our symmetry with an ``ugly'' Hamiltonian -the
realistic nucleon-nucleon interaction. This is in contrast to the more
prevalent practice of constructing simplified Hamiltonians  to display
approximate symmetries.

(b) We obtain a simpler result (equality of spin and orbit)
for the energy independent sum then for the energy weighted sum. In
most other cases, the energy-weighted strength gives the simplest
results.

(c) We even go beyond $2$ $\hbar\omega$ and show that
although the symmetry no longer holds, there is a wide plateau where
the cumulative spin and orbit sums are nearly equal.

We obtain this symmetry not despite of but because of the
fact that we remove spurious states. Interest in spurious states is
widespread -not only for nuclear structure but also in atomic physics
and for the structure of baryons where the degrees of freedom are
quarks and gluons. Thus the symmetry we have found here in the nuclear
context should be of interest in these other fields, even if only a
suggestion that something of interest may be lurking in the shadows.
And indeed even in the present context, it may suggest to others that
it is worth probing more deeply for unexpected symmetries.

We thank Alex Brown and Mihai Horoi for useful communications. This
work was supported by U.S. Department of Energy under Grant
DE-FG05-86ER-40299.

\begin{figure}
\caption{Sum of B(M1)$_{\rm spin}$ (solid line) and of B(M1)$_{\rm
orbit}$ (dashed line) with $x$=1, $y$=1 and up to 6 $\hbar \omega$
admixtures in units of 10$^{-3}\mu_{N}^{2}$.}
\end{figure}

\begin{table}
\caption{Summed spin and orbital magnetic dipole moment strength in
$^4\mbox{He}$ in units of $10^{-3}{\mu_N}^2$.}
\begin{tabular}{cccccccc}
\multicolumn{2}{c}{Interaction}
& \multicolumn{2}{c}{up to $2$$\hbar \omega$}
& \multicolumn{2}{c}{up to $4$$\hbar \omega$}
& \multicolumn{2}{c}{up to $6$$\hbar \omega$} \\
\tableline
x & y & SPIN & ORBIT & SPIN & ORBIT & SPIN & ORBIT\\
%\hline
0 &0  &0.8546 & 0.8546 & 1.3357& 5.1635 & 1.5897 & 7.1474\\

1 &0  &0.8569 & 0.8571 & 1.3417 & 5.1851 & 1.6211 & 7.2296\\

0 & 1 &3.8245  & 3.8239  & 5.2346 & 10.937 & 6.0653 & 14.607\\

1 & 1 &3.3944  & 3.3955  & 4.8288 & 10.554 & 5.6052 & 14.272\\
\end{tabular}
\end{table}

\begin{table}
\caption{ For the case $x$=$0$ $y$=$0$ (central interaction
-$LS$ limit), we give the energies and $B(M1)$'s of `spin excited'
and `orbit excited' states, with up to $2$ $\hbar\omega$ admixtures.}
\begin{tabular}{cccc}
{}&{Energy} & \multicolumn{2}{c}{$B(M1)$} \\
{}&{(MeV)} & \multicolumn{2}{c}{(in units of $10^{-3}{\mu_N}^2$)}\\
\hline
& & SPIN & ORBIT \\
\hline
NON-SPURIOUS &36.7 & 0 &  0 \\
             &44.0 & 0 &  0 \\
             &45.3 & 0.855 &  0 \\
             &48.8 & 0 &  0 \\
             &49.2 & 0 &  0 \\
             &53.9 & 0 &  0 \\
             &56.5 & 0 &  0.855 \\
\hline
SPURIOUS     &436.7 & 0 &  0 \\
             &436.7 & 0 &  0 \\
             &436.7 & 0 &  0 \\
             &439.3 & 0 &  13.07 \\
\end{tabular}
\end{table}

\begin{table}
\caption{Summed {\underline {low energy}} spin and orbital
magnetic dipole moment strength in $^4\mbox{He}$ with up to $6$$\hbar
\omega$ admixtures, in units of $10^{-3}{\mu_N}^2$.}
\begin{tabular}{ccccc}
\multicolumn{2}{c}{Interaction}
& \multicolumn{2}{c}{$B(M1)$}
& \multicolumn{1}{c}{Deviation} \\
\tableline
x & y & SPIN & ORBIT & ($\%$)\\

0 &0  &0.5592 & 0.6018 & 7.3\\

1 &0  &0.5888 & 0.6291 & 6.6\\

0 & 1 &1.9546  & 1.8046 & -8.0\\

1 & 1 &1.8199  & 1.6155 & -10.8\\
\end{tabular}
\end{table}

\begin{table}
\caption{For the case $x$=$0$ $y$=$0$ (central interaction), we
give the energies and $B(M1)$'s of `spin excited' and `orbit excited'
states, with strength $\ge$$10^{-4}{\mu_N}^2$.}
\begin{tabular}{ccc}
{Energy} & \multicolumn{2}{c}{$B(M1)$} \\
{(MeV)} & \multicolumn{2}{c}{(in units of $10^{-3}{\mu_N}^2$)}\\
\tableline
 & SPIN & ORBIT \\

34.369 & 0.5524 & 0       \\
43.509 & 0       & 0.6015 \\
61.414 & 0       & 2.7900 \\
65.253 & 0.2048 &  0      \\
66.616 & 0.2276 &  0      \\
67.762 & 0.2265 &  0      \\
71.837 & 0.2334 &  0      \\
71.868 & 0       & 0.7696 \\
73.723 & 0       & 0.9369 \\
83.071 & 0       & 0.7107 \\
96.715 & 0       & 0.7761 \\
101.20 & 0       & 0.2435 \\
107.35 & 0       & 0.1216 \\

\end{tabular}
\end{table}
\end{document}